\documentclass[prl,twocolumn,showpacs]{revtex4}
\usepackage{graphicx}
 \usepackage{amsmath}
\usepackage{color}

\newcommand{\im}[1]{\mathrm{Im}\left.#1\right.}
\newcommand{\kzm}{k_{z}}

\newcommand{\cutoff}{\Lambda}
\renewcommand{\imath}[0]{\mathrm{i}}

\begin{document}

\title{Casimir interaction from magnetically coupled eddy currents}
\author{Francesco Intravaia}%
\author{Carsten Henkel }
\affiliation{Institut f\"{u}r Physik und Astronomie, Universit\"{a}t Potsdam, 
    14476 Potsdam, Germany
}
\date{06 Sep 2009}

\begin{abstract}
We study the quantum and thermal fluctuations of eddy (Foucault) currents 
in thick metallic plates. A Casimir interaction between two plates arises
from the coupling via quasi-static magnetic fields. As a function of distance,
the relevant eddy current modes cross over from a quantum to a thermal 
regime. These modes alone reproduce previously discussed thermal anomalies 
of the electromagnetic Casimir interaction between good conductors.
In particular, they provide a physical picture for 
the Casimir entropy whose nonzero value at zero temperature arises from 
a correlated, glassy state.
\end{abstract}


\maketitle



Spatially diffusive transport is a basic physical phenomenon that has
been studied with a wealth of methods.
For example, the equation for heat conduction was solved by J.\ Fourier 
by his transformation that provided later the framework for 
quantum field theory.
Remarkably, the Fourier modes of the diffusion 
equation itself,
\begin{equation}
	\partial_{t} A = D \nabla^2 A, 
    \label{eq:diffusion-equation}
\end{equation}
do not fit into a simple quantum field theory because they have
purely imaginary frequencies. The quantization of 
(over)damped modes
can be done, however, with an alternative approach, the `system+bath'
paradigm of dissipative quantum dynamics.  In this picture,
the observables relevant to a mode (specified, e.g., in Fourier space
by its wavevector ${\bf q}$) 
are damped in time because they strongly couple to a system with
infinitely many degrees of freedom that are not directly
accessible (`bath')\cite{Weiss08}. 
The additional fluctuations the bath couples into the system, 
compensate at equilibrium for dissipative loss\cite{Ford93a}, even
at zero temperature. This implies that a quantity 
like the zero-point energy of a damped mode is no longer given by 
the usual $\frac12 \hbar \omega_{\bf q}$ and must be redefined and 
reinterpreted. Indeed, the ground state of the combined system+bath 
is, in general, \emph{entangled}: the corresponding interaction
Hamiltonian is responsible for the change in the zero-point energy 
relative to the decoupled system\cite{Nagaev02,OConnell06}. 
Clearly this has an impact on all phenomena connected with 
fluctuation energy, for which the paradigmatic example is 
the Casimir effect\cite{Casimir48}.

In this letter we discuss a specific example of `diffusive modes': 
eddy (Foucault) currents in two metallic plates\cite{Jackson75}.  
We address their quantum 
and thermal fluctuations and their role in the electromagnetic 
Casimir interaction\cite{Casimir48}. Similar to the approach of
Ref.\cite{plasmons}, 
we isolate the
contribution of eddy currents among all other modes. We show that 
these modes quantitatively reproduce the ``unusual features'' of the 
Casimir force between good conductors at finite temperature (as 
predicted by Lifshitz theory), that have been intensely debated
for several years\cite{Milton04,Mostepanenko06}.  


To begin with, consider a metallic bulk 
described by a local dielectric function in Drude form:
$\varepsilon( \omega ) = 1 - \Omega^2 / [\omega(\omega + {\rm i}Ê
\gamma)]$ ($\Omega$: plasma frequency, $\gamma$: damping rate). 
This system allows for a continuum of damped, chargeless modes with a
(transverse) current density and (dominantly) magnetic fields at frequencies
of order $\gamma$ or below: they are called `eddy' or `Foucault
currents' and are at the base of phenomena like magnetic braking 
or the familiar induction oven\cite{Jackson75}. The `bath' for these 
modes is 
provided by the phonons in the metal or by impurity scattering.
On time scales slower than 
$1/\gamma$, the vector potential satisfies
Eq.(\ref{eq:diffusion-equation}) with an electromagnetic diffusion
constant $D = \gamma \lambda^2$ 
($\lambda \equiv c/\Omega$: plasma penetration depth, 
$\lambda \approx 20\,{\rm nm}$ for gold).
A mode analysis in Fourier space yields imaginary
frequencies $\omega_{\bf q} = - {\rm i} \xi_{\bf q} 
\approx - {\rm i} D q^2$ for ${q} \ll 1/\lambda$: 
due to the coupling to the electromagnetic field, the
damping of eddy currents becomes weaker (compared to 
$\gamma$) and dispersive. 

Let us now split the metal in two half-spaces, inserting a vacuum 
layer of thickness $L$ (``cavity'').
While the eddy currents cannot leave the medium, the associated 
electromagnetic 
fields do; they cross the cavity in the 
form of evanescent waves. 
The corresponding magnetic fields provide
Johnson noise that dominates, at low frequencies,
over the fluctuations of other 
modes confined within the 
cavity\cite{Torgerson04,Bimonte07a,Svetovoy07}.
The characteristic frequency 
is the so-called Thouless energy\cite{Thouless77}
\begin{equation}
    \hbar \xi_L \equiv \hbar D/L^2 
    \label{eq:def-diffusive-frequency}
\end{equation}
and the scattering rate $\gamma$ provides a 
strict upper limit to the $\xi_{\bf q}$. Typical numbers for gold at
room temperature
are $\hbar \xi_L \sim 20\,{\rm K}$ 
and $\hbar \gamma \sim 500\,{\rm K}$ at $L= 100\,{\rm nm}$.
We focus in the following on the polarization with the electric field
parallel to the metal surface (TE). Indeed, the orthogonal case (TM)
is associated with a surface charge that 
effectively decouples medium and vacuum.

\begin{figure}[tbph]
	\centering\includegraphics*[width=64mm]{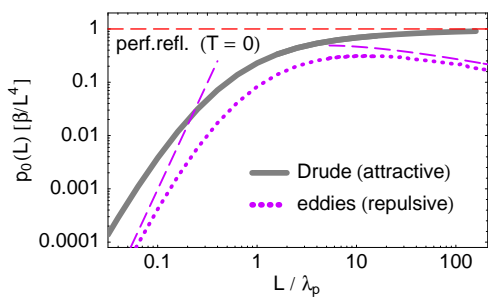}
\vspace*{-01mm}    
\caption[]{Casimir pressure ($T = 0$) between two metallic plates:
contribution of eddy currents alone and of all electromagnetic modes
(Drude metal) in the TE polarization. Normalization to the perfect 
reflector, $\pi^2\hbar c/(480 L^4)$,
distance $L$ in units of the the plasma wavelength
($\lambda_{\rm p} = 2\pi \lambda$).
Dashed lines: asymptotics~Eq.(\ref{eq:Casimir-short-large-dist}). 
We take $\gamma = 0.08\,\Omega$ and a cutoff $\Lambda = 5\,\gamma$.}
\label{fig:cas-pressures}
\end{figure}

For an overdamped mode,
the `system+bath' approach predicts a zero-point 
energy\cite{Nagaev02,Intravaia08}
\begin{equation}
    E_{\bf q} = -\frac{ \hbar \xi_{\bf q} }{ 2\pi } \log( \xi_{\bf q} / \cutoff)
    \label{eq:zpe-overdamped-mode}
\end{equation}
where $\Lambda$ is a cutoff frequency related to the bath response 
time\cite{Weiss08,Intravaia03}.
The corresponding Casimir energy captures the shift in the (imaginary) mode 
frequencies $\xi_q = \xi_q( L ) $ due to the magnetic cavity
field correlating the eddy currents in the two plates. We therefore sum this expression
over the eddy current mode continuum and subtract the reference system 
of two isolated half-spaces\cite{Casimir48}. 
The difference of mode density per area, $\tilde\rho( \xi; L )$ 
is related
to scattering phases at the metal-vacuum interface (see Appendix):
\begin{equation}
\label{eq:mu}
\tilde\rho(\xi; L) = -\partial_{\xi}\,
\int\!\frac{{\rm d}^2k}{\pi (2\pi)^{2}}
\im{ \log 
 \left[1 - r_k^{2}(-\imath 
 \xi + 0)e^{-2\kappa L}\right]
}
\end{equation}
where $\kappa=\sqrt{k^{2} + (\xi / c)^{2}}$ and $r_k$ the TE reflection coefficient. 
This shows power laws in $\xi$
below and above the Thouless frequency $\xi_L$, providing
the asymptotic behaviours [Fig.\ref{fig:cas-pressures}]
\begin{equation}
E_{D}( L ) \sim
\left\{
\begin{array}{ll}\displaystyle
\mbox{const.}
-\hbar \gamma L \lambda^{-3}
\log( \cutoff/\gamma  )
, & L\ll \lambda ,
\\[0.8em]
\displaystyle
\hbar (\xi_L c / L)^{1/2} L^{-2} 
\log( \cutoff L / c )
,  & L\gg c / \gamma
.
\end{array}
\right.
	\label{eq:Casimir-short-large-dist}
\end{equation}
%
%
The associated pressure $- {\rm d} E_D / {\rm d}L$ shows
a repulsive Casimir effect\cite{Torgerson04,Bimonte07a,Svetovoy07},
provided the cutoff is sufficiently large (a few times $\gamma$).
The repulsive character can be understood from the Lenz rule 
for current-current (diamagnetic) interactions\cite{Jackson75}.
Eq.(\ref{eq:Casimir-short-large-dist})
does not show the scaling $\hbar c / L^3$ characteristic of perfect
reflectors\cite{Casimir48} and is much weaker
because of the finite bandwidth $\lesssim \gamma$ of the relevant
mode continuum. The pressure vanishes for $\gamma\to 0$ because 
dynamical fluctuations induced by the bath are suppressed.
At small distances, $E_{\rm D}(L)$ essentially scales 
with the
`missing volume' $L \times \mbox{(plate area)}$ and the bulk mode
density $\sim \gamma^{1/2} D^{-3/2}$ (at $\xi \sim \gamma$)
of the eddy current continuum.
Their zero-point fluctuations exert, in this regime, 
a constant kinetic and magnetic pressure. 
\begin{figure}[tbp]
    \centering
    \includegraphics*[width=64mm]{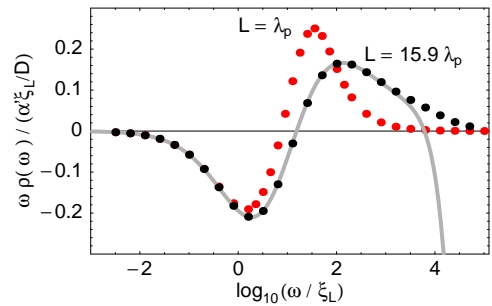}
\vspace*{-01mm}    
\caption[]{%
Eddy current mode density $\rho( \omega; L )$ 
for two plates at distance $L$
($\gamma = 10^{-3}\,\Omega$, 
$\lambda_{\rm p} = 2\pi \lambda$,
$\alpha'/L^2 = \xi_L |\rho(0; L)|$ 
[see Eq.(\ref{eq:estimate-low-frequency-DOS})]).
Frequency normalized to the Thouless frequency 
$\xi_{L} = D / L^2$.
Thick gray line: mode density used
in the Lifshitz theory with the Drude dielectric function. 
TE polarization only.} 
\label{fig:diffusive-dos}
\end{figure}



\begin{figure}[btph]
\centering 
\includegraphics*[width=64mm]{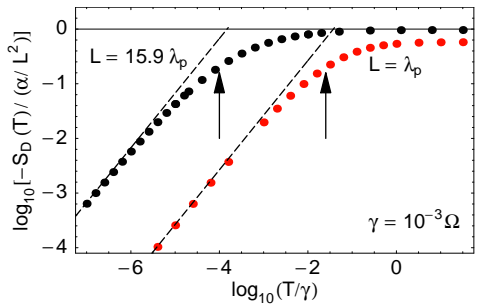}
\vspace*{-01mm}    
\caption{Eddy current Casimir entropy vs temperature $T$.
Dashed: Eq.(\ref{eq:entropy-regimes}) for $T \to 0$.
We normalize to the perfect reflector entropy
at high temperatures,
$\alpha / L^2 = \zeta( 3 ) k_{B}/ (16\pi L^2)$.
Arrows: crossover temperature 
$k_BT = \hbar\xi_{L} = \hbar D / L^2$.}
\label{fig:entropy}
\end{figure}

Let us now raise the temperature. 
Eddy currents are thermally excited at quite low temperatures
because of their low characteristic frequencies:
for good conductors at room temperature we have $\hbar \xi_L 
\ll k_B T < \hbar\gamma$ already at $L = 100\,{\rm nm}$. 
We find indeed that they give a significant contribution to the Casimir 
interaction. Consider first the 
Casimir entropy.
The contribution due to eddy currents can be calculated
in terms of a real frequency mode density (`DOS') $\rho( \omega; L )$ 
[Fig.\ref{fig:diffusive-dos}] using the following Kramers-Kronig like relation:
\begin{equation}
\rho( \omega; L ) 
= \int\!\frac{ {\rm d}\xi}{ \pi }\, \tilde{\rho}( \xi; L ) \frac{ \xi }{\xi^2 
+ \omega^2}.
\label{eq:KK-for-rho-of-omega}
\end{equation}
Fig.\ref{fig:diffusive-dos} shows that the DOS is significant in the frequency 
range $\omega \sim \xi_{L} \ldots \gamma$.
One also sees that in this range, the DOS of the 
electromagnetic Casimir effect between Drude metals (gray line) is entirely 
given by Foucault currents.
The zero frequency limit is
\begin{equation}
\rho( \omega \to 0; L ) \approx 
- \frac{2 \ln 2 - 1}{ 8 \pi^2 D } <0.
\label{eq:estimate-low-frequency-DOS}
\end{equation}
Given this DOS, we get the free energy $\mathcal{F}(T, L)$ by
multiplying with 
$k_B T\log[ 2\sinh(\hbar\omega/(2 k_B T))]$ and integrating
over $\omega$ (see Appendix).
The entropy, 
$S(T, L) = - \partial \mathcal{F} 
/ \partial T$, is cutoff independent
and becomes [see also Fig.\ref{fig:entropy}]
%
\begin{equation}
\frac{ S }{ k_{B} } = \left\{
\begin{array}{ll}\displaystyle
\displaystyle
 \frac{\pi^2}{3} \frac{ k_{B} T }{ \hbar } \rho( 0; L ) 
,	     & k_{B}T \ll \hbar \xi_L,
\\[0.85em]
\displaystyle
- \frac{ \zeta( 3 ) f(L/\lambda) }{ 16 \pi L^2 } = S_{\infty}(L)
,            & k_{B}T \gg \hbar \xi_{L},
\end{array}\right.
	\label{eq:entropy-regimes}
\end{equation}
where the dimensionless function $f(L/\lambda) \to 1$ for $L \gg 
\lambda$.
The characteristic temperature scale $\hbar \xi_{L} / k_{B}$
has been noticed previously from the Lifshitz theory 
with a temperature-independent scattering 
rate~\cite{Milton04,Torgerson04}.
A negative entropy is not surprising here since we are considering a 
difference, with the entropy for infinitely separated plates subtracted.
The sign means that one plate is 
acquiring information (entanglement) about the configuration in the other 
one through the electromagnetic interaction.
The entropy vanishes linearly as $T \to 0$, in agreement with the Nernst
heat theorem (third law of thermodynamics). In this limit, modes with 
frequencies above $k_{B}T/\hbar$ `freeze' to their ground state, do 
not contribute to the entropy, and a unique ground state for the system 
remains at $T = 0$. 
We have checked that the free energies at low $T$ due to eddy currents
alone and in the full Lifshitz theory\cite{Hoye07a,Ingold09} coincide in
their first two terms ($T^2$ and $T^{5/2}$).   
%

Another scenario emerges with a $T$-dependent scattering rate
when the ratio $\xi_L(T) / T$ vanishes as $T \to 0$.
This happens for example in a `perfect crystal'\cite{Bezerra04} 
where $\gamma( T ) \sim T^n$ ($n \ge 2$).  
In this case the Foucault modes stay in the high-temperature
regime and give a universal contribution
$S_{\infty}(L) \sim - f(L/\lambda) / 
L^2$ to the entropy. The same `entropy defect' 
has been found in previous work on the Lifshitz theory 
with the Drude model\cite{Bezerra04}
and traced back to evanescent TE modes\cite{Svetovoy07}. 
Nevertheless this does not conflict with thermodynamics because, in this limit, the ground state 
becomes infinitely degenerate: for $D \to 0$,  
the DOS $\rho(0;L) \to \infty$ 
in Eq.(\ref{eq:estimate-low-frequency-DOS}),
and the spectrum collapses to zero frequency
because the characteristic frequency scale $\xi_L \to 0$.
The diffusive modes then become 
static current loops interlocking with 
magnetic field lines that permeate the metallic bulk, similar to the 
frozen magnetization in an ideally conducting medium.
This is a glassy state with a nonzero entropy (`Foucault glass').
As a check of this idea we have formulated a Hamiltonian field theory 
for a glass of static currents and have calculated the
change in entropy per area for two magnetically coupled half-spaces,
recovering a result that perfectly coincides with $S_{\infty}( L )$.
It is clear, of course, that the Foucault glass is the result of an
over-idealized characterization (`perfect crystal').
Indeed, the Lifshitz entropy vanishes with $T\to 0$ when additional, 
even small, relaxation mechanisms like 
Landau damping are included\cite{Sernelius05aSvetovoy06a}.

\begin{figure}[btph]
    \centering \includegraphics*[width=64mm]{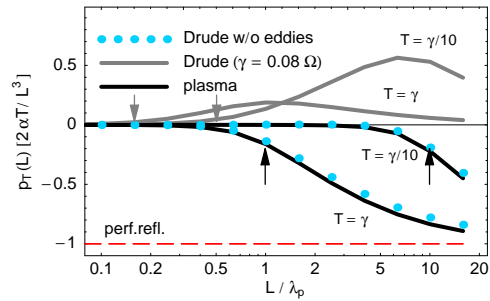}
\vspace*{-01mm}    
\caption[]{Thermal Casimir pressures
    (TE polarization only)
    vs.\ distance $L$, normalized to the perfect reflector
[$\zeta(3) k_B T / (8\pi L^3)$].
In the Drude model (gray, $\gamma = 0.08\,\Omega$), the 
pressure is repulsive and vanishes for
    $L \gg \hbar c / (2 k_B T)$ (black arrows).
    Blue dots: Drude model with eddy current pressure
subtracted.
Black: plasma model.
    Gray arrows: electromagnetic diffusion length 
    $[\hbar D/ (k_B T)]^{1/2}$.
}
    \label{fig:Casimir-pressure}
\end{figure}

Let us now consider the eddy current contribution to the 
Casimir pressure at $T > 0$.
The debate around
the electromagnetic Casimir interaction\cite{Milton04,Mostepanenko06}
has revealed 
that the large-distance pressure between Drude metals
is effectively halved compared to perfectly reflecting mirrors, even for an 
infinitesimally small dissipation rate. This is
illustrated in Fig.\ref{fig:Casimir-pressure} where the 
temperature-dependent
part of the Casimir pressure 
is plotted for the Drude model (gray lines vanishing as $L \gg 
\lambda_T \equiv \hbar c / (2 k_B T)$). 
The Drude model thus cannot describe the limit
of a perfect reflector (dashed line in 
Fig.\ref{fig:Casimir-pressure}), even if we take 
$\gamma \to 0$ or $\lambda \to 0$
(`ideal conductor'). 
Models that go over to the perfect reflector (as $L \gg \lambda_T$) 
are superconductors\cite{Bimonte08a,Haakh09ip} 
and the so-called `plasma model' (lossless dielectric function
with $\gamma = 0$ right from the start, solid line
in Fig.\ref{fig:Casimir-pressure}). 
The thermal pressure from eddy currents accounts nearly 
completely for the difference between the Drude and the previous models \cite{Torgerson04,Bimonte07a,Svetovoy07}: they thus represent very precisely
the object under debate. In particular eddy currents provide a simple physical explanation
why in the Drude model, the thermal Casimir pressure sets in 
at unusually short distances (compared to the thermal wavelength, 
arrows in Fig.\ref{fig:Casimir-pressure}):
this happens indeed when the dissipative energy scale crosses over from the
quantum into the classical regime, $\hbar D / L^2 
\sim k_B T$
\cite{Ingold09}. 
The absence of eddy currents in the plasma model is actually not 
surprising since it behaves like a superconductor and expels 
low-frequency magnetic fields\cite{London35}. 
The previously discussed `perfect crystal' limit ($D \to 0$ of the Drude model), 
on the contrary, corresponds to the
ideal conductor that does not show the Mei\ss ner effect\cite{London35}.

Recent measurements of the Casimir interaction between 
gold-coated bodies (a plate and a sphere) 
at room temperature have been argued to favor a theory within the
(lossless) plasma model\cite{Decca05}, but these results do not 
appear to be commonly accepted yet. 
In view of our analysis, a similar 
prediction could equally emerge within the Drude model if the 
contribution of 
eddy currents were somehow reduced, keeping only the other (photonic)
modes. How this should be
implemented physically, remains to be understood in detail. Potential 
mechanisms should take into account, however, that in other contexts,
thermally excited eddy currents have been observed, in agreement with
the usual thermodynamic theory\cite{Kittel05}.

To reduce the eddy current Casimir interaction, we have 
considered a calculation at constant entropy, noting 
that also the thermalization rate of low-frequency 
modes is very small. This approach leads, however, to a significant 
overshooting of the Casimir pressure at large distances (very low 
specific heat according to Eq.(\ref{eq:entropy-regimes})). 
Any mechanism that gives a nonzero real part to the 
frequencies of eddy current modes (see e.g.\ Ref.\cite{Bimonte07a}), 
would reduce their 
contribution at low temperatures because they then `freeze out' 
at $k_B T \ll {\rm Re}\,\hbar \omega_k$.
This is not likely for current room temperature experiments, however, 
as estimates based on finite-size effects show.

In summary, we have discussed an example of quantum thermodynamics
for an overdamped field, calculating the Casimir energy due to (suitably
generalized) zero-point energies and its dependence on temperature. 
This applies to a simple physical system of two metallic plates that has
been controversially discussed for several years by now. We have 
considered diffusive (eddy or Foucault) currents in normally conducting
plates and their correlations via magnetic fields that lead to a repulsive
pressure. These modes have a
range of characteristic frequencies that is very different from photon 
modes. They cross over from a quantum to a thermal regime in a range
of parameters that is accessible with current experiments, namely when
the energy scale of diffusive transport (Thouless energy $\hbar D / L^2$) 
exceeds $k_B T$. The finite-temperature 
%
quantum field theory of these modes gives significant entropic corrections to 
the Casimir pressure that set in 
at distances shorter than the thermal photon wavelength and 
explain quantitatively
the large thermal corrections to the Casimir pressure in the Lifshitz theory 
for Drude metals.  
We reproduce the
nonzero Casimir entropy for this system and suggest a physical understanding
of its unusual behaviour. In the particular case that the damping
rate becomes infinitesimally small at low $T$, an entropy defect
remains that has been interpreted in terms of a 
glassy state of quasi-static Foucault currents, the Nernst theorem being not
applicable. This limit does not recover
the lossless case (plasma model) because there, magnetic fields 
are expelled from the bulk, as in the London
description of a superconductor. 
Similar conclusions could also apply to other classes of modes (like overdamped coupled
surface plasmons\cite{plasmons}) providing insight into similar controversies \cite{Chen07b-Svetovoy08b} and perhaps a route to engineer the Casimir force, similar to the suggestion in\cite{plasmons}. 
From a theoretical point of view, our analysis illustrates that the unusual 
behavior of quantum field theory for normally conducting mirrors finds a clear 
physical justification in the framework of open quantum systems, glassy 
dynamics, and superconductivity, and paves the way for further theoretical and experimental investigations.

\paragraph{Acknowledgments. ---}
The work of FI was supported by the Alexander
von Humboldt Foundation and in part by the US National Science 
Foundation
(Grant No.\  PHYS05-51164).
 We acknowledge funding
from the ESF research network programme
``New Trends and Applications of the Casimir Effect'' and the Deutsche 
Forschungsgemeinschaft.
We thank G. Bimonte, D. Dalvit, S. Ellingsen, J.-J. Greffet, H. 
Haakh, B. Horovitz, G.-L. Ingold, G. Klimchitskaya,
A. Lambrecht, S. K. Lamoreaux, 
V. M. Mostepanenko,
L. Pitaevskii, S. Reynaud, F. da Rosa, and F. Sols for instructive 
comments.

\subsection{Appendix: eddy current mode density}


For the density of modes at real frequencies, $\rho( \omega; L)$,
of the electromagnetic Casimir effect we use the following prescription: 
the free energy (per unit area) is written in the form 
\begin{equation}
	F = \int\limits_0^{\infty}\!{\rm d}\omega\, \rho( \omega; L)
	f( \omega )
	\label{eq:free-Casimir-energy}
\end{equation}
where $f( \omega ) = k_B T \log[2 \sinh(\hbar \omega / 2 k_B T)]$ 
is the free energy per mode.
For two parallel, identical plates, this gives
\begin{eqnarray}
	&&\rho( \omega; L ) = - \frac{ 1 }{ \pi } \partial_{\omega}
	{\rm Im}\, 
	\sum_{\mu }
	\int\frac{ {\rm d}^2 k }{ (2\pi)^2 }
	\log D_\mu( {\bf k}, \omega + {\rm i} 0 )
	\label{eq:def-rho-of-omega}
	\\
	&& D_\mu( {\bf k}, \omega) = 1 - r^{2}_\mu( k, \omega)\, 
	{\rm e}^{-2\kappa L}
	\label{eq:def-dispersion-D}
\end{eqnarray}
The integration is over the wavevector ${\bf k}$ parallel to the plates,
and the two polarizations $\mu  = {\rm TE},\, {\rm TM}$
are summed over. The $r_\mu( k, \omega)$
are the reflection coefficients for the electromagnetic field from a single
half-space, and $\kappa$ is defined after Eq.(\ref{eq:TETMcoeff}) below.
Passivity does not allow for poles and zeros of the dispersion function 
$D_\mu({\bf k},\omega)$ in the upper half of the complex plane. Its 
analytical continuation into the lower half plane does show, however, poles,
zeros and branch cut singularities: the zeros (poles) actually define the
complex mode frequencies of the two mirror system (of two isolated mirrors),
respectively.
Branch cuts arise by taking a difference of mode densities in the 
continuous part of the spectrum. The expansion of the mode density
$\rho( \omega; L)$ over these singularities, displaying explicitly the contribution
of a branch cut along the negative imaginary axis, reads\cite{Intravaia08}
\begin{equation}
\label{eq:poleExpansion}
\rho(\omega; L) = - \int\frac{ {\rm d}\xi }{\pi} {\rm Im}
\frac{\tilde\rho( \xi; L ) }{\omega + \imath\xi }
+ \rho_{\rm other\,modes}( \omega; L)
\end{equation}
where we have introduced the (real-valued) mode density along the branch 
cut $\tilde\rho( \xi; L)$. The first term is Eq.(6) of the main paper.

In order to calculate the mode density $\tilde\rho( \xi; L)$, we 
specialize to a specific form for 
the reflection coefficients: namely the Fresnel form with the Drude 
model
\begin{equation}
\label{eq:Drude}
\varepsilon( \omega ) = 1-\frac{\Omega^{2}}{\omega(\omega+\imath\gamma)}
\end{equation}
involving the plasma frequency $\Omega$ and $\gamma$ the
scattering rate for the current density. 
In the TE and TM polarization, respectively \cite{Jackson75}
\begin{equation}
\label{eq:TETMcoeff}
r_{TE}(k,\omega)=\frac{\kappa-\kappa_{m}}{\kappa+\kappa_{m}}, 
\quad 
r_{TM}(k,\omega)
=
\frac{\varepsilon( \omega )\kappa-\kappa_{m} 
	}{ \varepsilon( \omega )\kappa+\kappa_{m} }
\end{equation} 
where $\kappa=\sqrt{k^{2} - (\omega / c)^{2}}$ and 
$\kappa_{m}=\sqrt{k^{2} - (\omega / c)^{2}\varepsilon( \omega )}$.
The branch cut at negative imaginary frequencies $\omega = - {\rm i}\xi$
is associated with the square root $\kappa_{m}$; it connects the solutions
of the equations
\begin{equation}
\label{eq:}
k^{2} - \varepsilon( \omega )(\omega / c)^{2}=0, \, -\infty
\end{equation}
The second equation easily gives $\omega=-\imath \gamma$;
the first equation is a third order equation of which we pick the purely
imaginary root $\omega = - {\rm i}\xi_{\bf k}$.
(The other two correspond to the plasma edge above which electromagnetic
waves penetrate into the bulk of the mirrors.)
In the limit $k\ll \Omega/c$, we get 
\begin{equation}
\label{eq:branchpoint}
\omega = -\imath\xi_{\bf k}\approx 
-\imath D {k}^{2}
\end{equation} 
where $D = \gamma c^2 / \Omega^2$ 
is the diffusion coefficient. Both reflection coefficients 
and hence the dispersion function~(\ref{eq:def-dispersion-D})
show a branch cut for
$\omega \in [-\imath\xi_{\bf k},-\imath\gamma]$.
The mode density $\tilde\rho( \xi; L)$ along this cut is found 
from the argument principle, taking an integration contour around the 
negative imaginary axis. This gives
\begin{equation}
\label{eq:mu}
\tilde\rho(\xi; L) = -\partial_{\xi} {\rm Im}\,
\sum_{\mu}
\int\!\frac{{\rm d}^2k}{\pi (2\pi)^{2}}
\im{ \log D_\mu( {\bf k}, -\imath \xi + 0 )
}
\end{equation}
where the $+ 0$ prescribes the side of the branch cut in the right 
half plane (setting the sign of the wave vector $\kappa_m$).

The same result can also be obtained using a scattering approach.
In this case, we begin with the dispersion relation in a bulk Drude 
metal, 
\begin{equation}
    0 =
    \frac{ c^2    {\bf q}^2 - \omega^2  \varepsilon( \omega ) }{
    c^2    {\bf q}^2 - \omega^2 }
    = 
    1
    + \frac{ \Omega^2 }{ c^2 {\bf q}^2 - \omega^2 }
    \frac{ \omega }{ \omega + {\rm i} \gamma } 
    \label{eq:bulk-modes-Drude}
\end{equation}
where $\varepsilon( \omega )$ is again the Drude model. 
Both dispersion functions in the numerator and denominator 
of this expression are a simple re-writing of the wave 
equations after Fourier transformation. 
The poles of Eq.(\ref{eq:bulk-modes-Drude})
illustrate the subtraction operated in the bulk dispersion
energy: the free space photon continuum and locally damped 
currents (not coupled to the long-range electromagnetic field).
Looking for the solutions to Eq.(\ref{eq:bulk-modes-Drude}),
three modes are found, of which one always occurs at imaginary
frequencies $\omega_{\bf q} = - {\rm i} \xi_{\bf q}$.  Its limits for small
and large $q$ are
\begin{eqnarray}
    q \ll \Omega/c : &  & \omega_{\bf q} \approx - {\rm i} D {q}^2
    \label{eq:diffusive-limit}  \\
    q \gg \Omega/c : &  & \omega_{\bf q} \approx - {\rm i} \gamma
    \label{eq:overdamped-limit}
\end{eqnarray}
where $D$ is again the electromagnetic diffusion constant.
In view of planar surfaces for the metal, we label the modes by the 
wavevector ${\bf k}$ projected onto the plane $z = 0$ and the
frequency $- {\rm i}\xi$. The perpendicular wavevector 
$\kzm$ is then fixed by the dispersion relation~(\ref{eq:bulk-modes-Drude}). 

The electric current density associated with diffusive waves
cannot escape into vacuum (except for the displacement current).  Hence 
we get total internal reflection with a coefficient of modulus unity.  
In the vacuum gap, the (transverse) electromagnetic field varies 
exponentially with a decay constant $\kappa^2 = {\bf k}^2 + \xi^2 / c^2$.  
The single-interface reflection coefficient is obtained applying the usual 
boundary conditions for the electromagnetic field. In the TE-polarization, 
we find
\begin{equation}
    r_{D}( {k}, - {\rm i}\xi ) =
        - \frac{ \kappa + {\rm i} \kzm }{ \kappa - {\rm i} \kzm } 
        =  - r_{TE}( {k}, - {\rm i}\xi + 0 ) 
    \label{eq:def-single-interface-reflection}
\end{equation}
With two interfaces at $z = \pm L/2$, odd and even modes
with respect to the mid-plane of the cavity vary as $\sinh \kappa z$
and $\cosh \kappa z$ for $|z| \le L /2$.  In the
medium, we find that the reflection coefficient $r_D$ is modified by a 
phase shift ${\rm e}^{-2 {\rm i}\delta }$, and this gives access to the 
change in the mode density (i.e., the DOS $\tilde\rho( {\bf k}, \xi )$) of 
the diffusive 
wave continuum~\cite{Barton79}. The phase shifts are measured with respect
to the configuration where the mirrors are placed an infinite distance
apart. Alternatively, we can compute the transmission coefficient
across the vacuum gap and get the DOS from its phase\cite{Bordag01}.
Summing over even and odd modes  we get a phase shift
  \begin{equation}
     {\rm e}^{ - 2 {\rm i} \delta } = 
     \frac{ 1 - (r_D^*)^{2} \,{\rm e}^{-2\kappa L} }{ 
     1 - r_D^{2} \,{\rm e}^{-2\kappa L} }
     \label{eq:recover-Lifshitz-form}
 \end{equation}
The mode density along imaginary frequencies, 
per $k$-vector and polarization, is thus
\begin{equation}
        \tilde\rho( {\bf k}, \xi ) = -\frac{ 1 }{ \pi }
	\frac{ \partial }{ \partial \xi }
        {\rm Im}\, \log\left[ 
        1 - r_D^{2}( k, - {\rm i} \xi) \,  {\rm e}^{-2\kappa L} \right]
    \label{eq:def-DOS-along-xi1}
\end{equation}
Integrating over the parallel wavevector and summing over the polarizations,
we recover Eq.(\ref{eq:mu}).











%
\end{document}